# Causality, Influence, and Computation in Possibly Disconnected Dynamic Networks[⋆]

Othon Michail[1,2], Ioannis Chatzigiannakis[1,2], and Paul G. Spirakis[1,2]

[1] Computer Technology Institute & Press "Diophantus" (CTI), Patras, Greece
[2] Computer Engineering and Informatics Department (CEID), University of Patras, Patras, Greece
Email: {michailo, ichatz, spirakis}@cti.gr

**Abstract.** In this work, we study the propagation of influence and computation in dynamic distributed systems. We focus on broadcasting models under a worst-case dynamicity assumption which have received much attention recently. We drop for the first time in worst-case dynamic networks the common *instantaneous connectivity* assumption and require a minimal *temporal connectivity*. Our temporal connectivity constraint only requires that *another causal influence occurs within every time-window of some given length*. We establish that there are dynamic graphs with *always disconnected instances* with equivalent temporal connectivity to those with always connected instances. We present a termination criterion and also establish the computational equivalence with instantaneous connectivity networks. We then consider another model of dynamic networks in which each node has an underlying communication neighborhood and the requirement is that each node covers its local neighborhood within any time-window of some given length. We discuss several properties and provide a protocol for counting, that is for determining the number of nodes in the network.

**Keywords:** dynamic network, distributed computing, mobile computing, worst-case dynamicity, influence, causality, temporal connectivity, termination, counting

---
[⋆] This work has in part been supported by the National Strategic Reference Framework (NSRF) (Regional Operational Programme - Western Greece) under the title "Advanced Systems and Services over Wireless and Mobile Networks" (number 312179) and by the Fourth Strategic Objective of the Operational Programme (OP) "Education and Lifelong Learning" (EdLL), entitled "Supporting the Human Capital in order to Promote Research and Innovation", under the project title "Foundations of Dynamic Distributed Computing Systems" (FOCUS).

# 1 Introduction

Distributed computing systems are more and more becoming dynamic. The static and relatively stable models of computation can no longer represent the plethora of recently established and rapidly emerging information and communication technologies. In recent years, we have seen a tremendous increase in the number of new mobile computing devices. Most of these devices are equipped with some sort of communication, sensing, and mobility capabilities. Even the Internet has become mobile. The design is now focused on complex collections of heterogeneous devices that should be robust, adaptive, and self-organizing, possibly moving around and serving requests that vary with time. Delay-tolerant networks are highly-dynamic, infrastructure-less networks whose essential characteristic is a possible absence of end-to-end communication routes at any instant. Mobility may be *active*, when the devices control and plan their mobility pattern (e.g. mobile robots), or *passive*, in opportunistic-mobility networks, where mobility stems from the mobility of the carries of the devices (e.g. humans carrying cell phones) or a combination of both (e.g. the devices have partial control over the mobility pattern, like for example when GPS devices provide route instructions to their carriers). Thus, it can vary from being completely predictable to being completely unpredictable. Gossip-based communication mechanisms, e-mail exchanges, peer-to-peer networks, and many other contemporary communication networks all assume or induce some sort of highly-dynamic communication network.

The formal study of dynamic communication networks is hardly a new area of research. There is a huge amount of work in distributed computing that deals with causes of dynamicity such as failures and changes in the topology that are rather slow and usually eventually stabilize (like, for example, in self-stabilizing systems [Dol00]). However the low rate of topological changes that is usually assumed there is unsuitable for reasoning about truly dynamic networks. Even graph-theoretic techniques need to be revisited: the suitable graph model is now that of a *dynamic graph* (a.k.a. *temporal graph* or *time-varying graph*) (see e.g. [KKK00,Kos09,CFQS11]), in which each edge has an associated set of time-labels indicating availability times. Even fundamental properties of classical graphs do not carry over to their temporal counterparts. For example, Kempe, Kleinberg, and Kumar [KKK00] found out that there is no analogue of Menger's theorem (see e.g. [Bol98] for a definition) for arbitrary temporal networks, which additionally renders the computation of the number of node-disjoint *s-t* paths **NP**-complete. Even the standard network diameter metric is no more suitable and has to be replaced by a dynamic/temporal version. In a dynamic star graph in which all leaf-nodes but one go to the center one after the other in a modular way, any message from the node that enters last the center to the node that never enters the center needs $n-1$ steps to be delivered; that is the *dynamic diameter* is $n-1$ while, one the other hand, the classical diameter is just 2 [AKL08] (see also [KO11] for a nice presentation).

As far as we are concerned, the two most outstanding models for dynamic distributed systems are the *population protocol* model of Angluin, Aspnes, Diamadi, Fischer, and Peralta [AAD$^+$06] and the *continuous connectivity* model of O'Dell and Wattenhofer [OW05] that further evolved to *T-interval connectivity* in the seminal STOC paper of Kuhn, Lynch, and Oshman [KLO10].

# 2 Related Work

In the population protocol model [AAD$^+$06], the computational agents are passively mobile, interact in ordered pairs, and the connectivity assumption is a *strong global fairness condition* according to which all events that may always occur, occur infinitely often. These assumptions give rise to some sort of structureless interacting automata model. The usually assumed *anonymity* and *uniformity* (i.e. $n$ is not known) of protocols only allow for commutative computations that eventualy stabilize to a desired configuration. Most computability issues in this area have now been established. Constant-state nodes on a complete interaction network (and several variations) compute the *semilinear predicates* [AAER07]. Semilinearity persists up to $o(\log \log n)$ local space but not more than this [CMN+11]. If constant-state nodes can additionally leave and update fixed-length pairwise marks then the computational power dramatically increases to the commutative subclass of **NSPACE**$(n^2)$ [MCS11a]. Other important works include [DGFGR06,GR09]. For a very recent introductory text see [MCS11b].

Distributed systems with worst-case dynamicity were first studied in [OW05]. Their outstanding novelty was to assume a communication network that may change arbitrarily from time to time subject to the condition that each instance of the network is connected. They studied asynchronous communication and considered nodes that can detect local neighborhood changes; these changes cannot happen faster than it takes for a message to transmit. They studied *flooding* (in which one node wants to disseminate one piece of information to all nodes) and *routing* (in which the information need only reach a particular destination node $t$) in this setting. They described a uniform protocol for flooding that terminates in $O(Tn^2)$ rounds using $O(\log n)$ bit storage and message overhead, where $T$ is the maximum time it takes to transmit a message. They conjectured that without identifiers (IDs) flooding is impossible to solve within the above resources. Finally, a uniform routing algorithm was provided that delivers to the destination in $O(Tn)$ rounds using $O(\log n)$ bit storage and message overhead.

Computation under worst-case dynamicity was further and extensively studied in a series of works by Kuhn *et al.* in the synchronous case. In [KLO10], the network was assumed to be *T-interval connected* meaning that any time-window of length $T$ has a static connected spanning subgraph (persisting throughout the window). Among others, *counting* (in which nodes must determine the size of the network) and *all-to-all token dissemination* (in which $n$ different pieces of information, called tokens, are handed out to the $n$ nodes of the network, each node being assigned one token, and all nodes must collect all $n$ tokens) were solved in $O(n^2/T)$ rounds using $O(\log n)$ bits per message, almost-linear-time randomized approximate counting was established for $T = 1$, and two lower bounds on token dissemination were given. In [KLLO10], it was shown that there is clock synchronization algorithm that achieves a clock difference of $O(D)$ between any pair of nodes at any time, and a clock difference of $O(\log D)$ between pairs of nodes that remain neighbors for $\Theta(D)$ rounds, where $D$ is the dynamic diameter. Several variants of *coordinated consensus* in 1-interval connected networks were studied in [KOM11]. Two interesting findings were that in the absence of a good initial upper bound on $n$, eventual consensus is as hard as computing deterministic functions of the input and that *simultaneous consensus* can never be achieved in less than $n - 1$ rounds in any execution. Requiring continuous connectivity has been supported by the findings of [CKLWL09], where a *connectivity service* for mobile robot swarms that encapsulates an arbitrary motion planner and can refine any plan to preserve connectivity while ensuring progress was proposed. [Hae11,HK11] are recent works that present information spreading algorithms in worst-case dynamic networks based on *network coding*. One of their important results is that the k-gossip problem on the adversarial model of [KLO10] can be solved using network coding in $O(n + k)$ rounds assuming the token sizes are sufficiently large ($\Omega(n \log n)$ bits). An *open* setting (modeled as high churn) in which nodes constantly join and leave has very recently been considered in [APRU12]. The focus was on robust and fast distributed computation in highly dynamic Peer-to-Peer (P2P) networks for the fundamental *distributed agreement* problem. For an excellent introduction to distributed computation under worst-case dynamicity see [KO11]. Two very thorough surveys on dynamic networks are [Sch02,CFQS11].

# 3  Contribution

In this work, we study worst-case dynamic networks that are *free of any connectivity assumption about their instances*. We only impose a *temporal connectivity* constraint stating that for each node $u$ and each time $t$ the state of node $u$ at time $t$ influences (see Section 4.1 for a formal definition of *causal influence*) at least one new node every $k$ steps (if an uninfluenced one exists). Note that temporal connectivity is the minimum assumption that allows for bounded end-to-end communication. We call this metric the *outgoing influence time* (oit) (similarly for the *incoming influence time*, or iit). In the 1-interval connectivity model the oit is 1, due to the fact that there is always an edge in the cut between the influenced and the uninfluenced nodes. We prove, in Section 4.2, that there are dynamic networks with oit 1 that are *disconnected in every instance*. To do this, while at the same time guaranteeing low edge-periodicity (with the edge-period of a dynamic graph defined as the fastest edge reappearence), we invoke a geometric edge-coloring method due to Soifer for coloring a complete graph of even order $n$ with $n - 1$ colors [Soi09].

Then, in Section 4.3, we prove that if a dynamic graph has unit oit then we cannot always have nodes influence at most one new node. Intuitively, this shows that information must necessarily spread faster than one new influence per node per step. Based on this and on the observation of some basic cases we conjecture that if the oit of a dynamic graph is 1 then there is always some node that only needs $\lfloor n/2 \rfloor$ steps to influence the whole network.

We also show (Section 4.4) that if nodes know that their oit is $k$ then they can infer an upper bound on their iit which provides them with the following *termination criterion*: a new incoming influence must occur in at most $kl(l + 1)/2$ steps, where $l$ is the number of existing incoming influences. We also present an example showing that when the oit is $k$ the iit may be up to $O(kn)$ and establish that any protocol for 1-interval connected graphs can be adapted to work on graphs with (outgoing or incoming) influence time $k$.

Then, in Section 5, we propose another model of worst-case temporal connectivity, called *local communication windows*, that assumes a *fixed underlying communication network* and restricts the adversary to allow communication between potential neighbors in every time-window of some fixed length. We prove some basic properties and provide a protocol for *counting* in this model. Formally, an algorithm is said to solve the counting problem if whenever it is executed in a dynamic graph comprising $n$ nodes, all nodes eventually terminate and output $n$.

Finally, in Section 6, we conclude and provide some interesting future research directions.

# 4  Spread of Influence in Dynamic Graphs

Let us begin with a few definitions. A *synchronous dynamic network* is modeled by a *dynamic graph* $G = (V, E)$, where $V$ is a static set of nodes and $E : \mathbb{N} \to \mathcal{P}(E')$, where $E' = \{\{u, v\} : u, v \in V\}$, (wherever we use $\mathbb{N}$ we mean $\mathbb{N}_{\geq 1}$) is a function mapping a round number $t \in \mathbb{N}$ to a set of undirected edges $E(t)$ drawn from $E'$. Intuitively, a dynamic graph $G$ is an infinite sequence of *instantaneous graphs* $G(t)$, $t \in \mathbb{N}$ ($G(t)$ denotes the instantaneous graph of $G$ at time $t$), whose edge sets are subsets of $E'$ chosen by a *worst-case adversary*. Throughout the paper, we denote $|V|$ by $n$. For some node $u$ and some time $t$, we denote by $(u, t)$ the state of $u$ at time $t$; the pair $(u, t)$ is called a *time-node*.

## 4.1  The Influence Time

Probably the most important notion associated with a dynamic graph is the *causal influence*. The causal influence captures the influence of the state of a node by the state of some other node and is the only means of capturing information spreading, an operation that is crucial for any distributed system.

**Definition 1 ([Lam78]).** *Given a dynamic graph $G = (V, E)$ we define an order $\to \subseteq (V \times \mathbb{N})^2$, where $(u, t) \to (v, t+1)$ iff $u = v$ or $\{u, v\} \in E(t+1)$. The* causal order *$\leadsto \subseteq (V \times \mathbb{N})^2$ is defined to be the reflexive and transitive closure of $\to$.*

Let us also define two very useful sets. Let $\text{past}_{(u,t')}(t) := \{v \in V : (v, t) \leadsto (u, t')\}$ and $\text{future}_{(u,t)}(t') := \{v \in V : (u, t) \leadsto (v, t')\}$, for $t \leq t'$. In words, $\text{past}_{(u,t')}(t)$ is the set of nodes whose $t$-state (i.e. their state at time $t$) has causally influenced the $t'$-state of $u$ and $\text{future}_{(u,t)}(t')$ is the set of nodes whose $t'$-state has been causally influenced by the $t$-state of $u$. Note that $v \in \text{past}_{(u,t')}(t)$ iff $u \in \text{future}_{(v,t)}(t')$.

Obviously, for a dynamic distributed system to operate as a whole there must exist some upper bound on the time needed for information to spread through the network. This is the weakest possible guarantee since without it global computation is impossible. The most abstract way to talk about information spreading is via the notion of the *dynamic diameter*. The *dynamic diameter* (also called *flooding time*) is a bound on the time required for each participant to causally influence and to be causally influenced by each other participant: formally, the dynamic diameter is the minimum $d$ s.t. for all times $t \in \mathbb{N}$ and all $u, v \in V$ it holds that $(u, t) \leadsto (v, t + d)$. A small dynamic diameter implies fast dissemination times. So, our first question is "how do graphs with small dynamic diameters look like"?

A class of dynamic graphs with small dynamic diameter is that of *$T$-interval connected* graphs [KLO10]. A dynamic graph $G = (V, E)$ is said to be $T$-interval connected for $T \geq 1$ if for all $r \in \mathbb{N}$, the static graph $G_{r,T} := (V, \bigcap_{i=r}^{r+T-1} E(r))$ is connected; that is, in any time-window of length $T$, a connected spanning subgraph is preserved. If $T$ is $\infty$ then a connected spanning subgraph must be preserved forever. It is not hard to see that the dynamic diameter of a $T$-interval connected graph is at most linear in $n$. In particular, the underlying connectivity ensures that for all $t \in \mathbb{N}$ the $t$-state of any node influences the $(t+n-1)$-state of any other node. Note that the propagation guarantee resulting from instantaneous connectivity is in fact more than a guarantee on the dynamic diameter. It ensures that *in each step* another node is causally influenced.

We now formalize the above novel influence metric. We define the *outgoing influence time* (oit) as the minimum $k \in \mathbb{N}$ s.t. for all $u \in V$ and all times $t, t' \in \mathbb{N}$ s.t. $t' \geq t$ it holds that

$$|\text{future}_{(u,t)}(t' + k)| \geq \min\{|\text{future}_{(u,t)}(t')| + 1, n\}.$$

Intuitively, the oit is the maximal time until the $t$-state of a node influences the state of another node (if an uninfluenced one exists) and captures the speed of information spreading. The *incoming influence time* (iit) is similarly defined as the minimum $k \in \mathbb{N}$ s.t. for all $u \in V$ and all times $t, t' \in \mathbb{N}$ s.t. $t' \geq t$ it holds that

$$|\text{past}_{(u,t'+k)}(t)| \geq \min\{|\text{past}_{(u,t')}(t)| + 1, n\}.$$

We can now say that the oit of a $T$-interval connected graph is 1 and that the iit can be up to $n - 2$. However, is it necessary for a dynamic graph to be $T$-interval connected in order to achieve unit oit? In the following section, we answer this question to the negative.

### 4.2 Dynamic Graphs with Disconnected Instances

We will show that there are always-disconnected graphs with optimal oit. We first present a simple example. Then we argue that its drawback is its high edge-periodicity: all edges reappear in only 2 steps. By invoking a geometric edge-coloring method we arrive at an always-disconnected graph with unit oit and low edge-periodicity (no edge reappears in less than $n - 1$ steps).

First, let us make a simple but useful remark:

**Proposition 1.** *If a dynamic graph $G = (V, E)$ has oit (or iit) 1 then each instance has at least $\lceil n/2 \rceil$ edges.*

*Proof.* $\forall u \in V$ and $\forall t \geq 1$ it must hold that $\{u, v\} \in E(t)$ for some $v$. In words, at any time $t$ each node must have at least one neighbor since otherwise it influences (or is influenced by) no node during round $t$. A minimal way to achieve this is by a perfect matching in the even-order case and by a matching between $n - 3$ nodes and a linear graph between the remaining 3 nodes in the odd-order case. □

Proposition 1 is easily generalized as: if a dynamic graph $G = (V, E)$ has oit (or iit) $k$ then for all times $t$ it holds that $|\bigcup_{i=t}^{t+k-1} E(i)| \geq \lceil n/2 \rceil$. Theorem 1 below will among others establish that this bound is tight.

It is not hard to see that the following is a simple dynamic graph with oit 1 and always disconnected instances: take a ring of $2l$ nodes, partition the edges into 2 perfect matchings $A$ and $B$ and alternate round after round between the edge sets $A$ and $B$. However, in this construction each edge reappears every second step. Intuitively, this dynamic graph may be conceived as simple as it has small *edge-period*. We now formalize this notion.

**Definition 2.** *The edge-period of a dynamic graph $G = (V, E)$ is defined as the minimum $p \in \mathbb{N}$ s.t., $\exists e \in \{\{u, v\} : u, v \in V\}$ and $\exists t \geq 1$, $e \in E(t) \cap E(t + p)$.*

Clearly, the edge-period of the dynamic graph described above is 2, because no edge ever reappears in 1 step and some, at some point, (in fact, all and always) reappears in 2 steps. We pose now an interesting question: Is there an always-disconnected dynamic graph with unit oit and edge-period $n - 1$? Note that this is harder to achieve as it allows of no edge to ever reappear in less than $n - 1$ steps.

To answer the above question, we define a very useful dynamic graph coming from the area of edge-coloring.

**Definition 3.** *We define the following dynamic graph $S$ based on an edge-coloring method due to Soifer [Soi09]: $V(S) = \{u_1, u_2, \ldots, u_n\}$ where $n = 2l$, $l \geq 2$. Place $u_n$ on the center and $u_1, \ldots, n_{n-1}$ on the vertices of a $(n - 1)$-sided polygon. For each time $t \geq 1$ make available only the edges $\{u_n, u_{m(0)}\}$ for $m(j) := (t - 1 + j \mod n - 1) + 1$ and $\{u_{m(-i)}, u_{m(i)}\}$ for $i = 1, \ldots, n/2 - 1$; that is make available one edge joining the center to a polygon-vertex and all edges perpendicular to it. (e.g. see Figure 1 for $n = 8$ and $t = 1, \ldots, 7$).*

In Soifer's dynamic graph, denote by $N_u(t) := i : \{u, u_i\} \in E(t)$, that is the index of the unique neighbor of $u$ at time $t$. The following lemma states that the next neighbor of a node is in almost all cases (apart from some trivial ones) the one that lies two positions clockwise from its current neighbor.

**Lemma 1.** *For all times $t \in \{1, 2, \ldots, n - 2\}$ and all $u_k$, $k \in \{1, 2, \ldots, n - 1\}$ it holds that $N_{u_k}(t + 1) = n$ if $N_{u_k}(t) = (k - 3 \mod n - 1) + 1$ else $N_{u_k}(t + 1) = (k + 1 \mod n - 1) + 1$ if $N_{u_k}(t) = n$ and $N_{u_k}(t + 1) = (N_{u_k}(t) + 1 \mod n - 1) + 1$ otherwise.*

**Theorem 1.** *For all $n = 2l$, $l \geq 2$, there is a dynamic graph of order $n$, with oit equal to 1, edge-period equal to $n - 1$, and in which each instance is a perfect matching.*

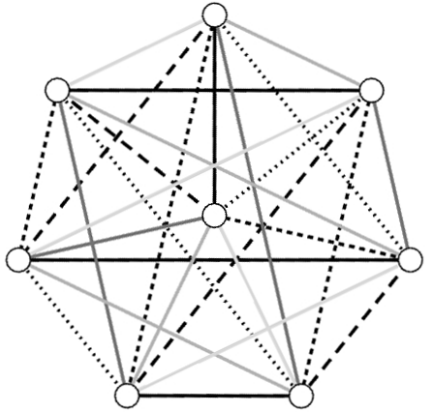

**Fig. 1.** Soifer's dynamic graph for $n = 8$ and $t = 1, \ldots, 7$.

*Proof.* Since $K_n$ is 1-factorable [3] it can be decomposed into $n - 1$ uniquely edge-colored (from $\{u_1, u_2, \ldots, u_{n-1}\}$) perfect matchings (we interpet each color as a time at which the corresponding edge is available). Consequently, $\forall u \in V$, the edges incident to $u$ are colored $\{1, 2, \ldots, n - 1\}$ and no such edge is adjacent to some other edge of the same color. Now extend this coloring as follows: each edge with color $r$ is available at times $r + b(n - 1)$ for $b \geq 0$.

We prove now that the **oit** is 1. We focus on the set $\text{future}_{(u_n, 0)}(t)$, that is the outgoing influence of the initial state of the node at the center. Note that symmetry guarantees that the same holds for all time-nodes (any node can be moved to the center without altering the graph). $u_n$ at time 1 meets $u_1$ and thus $\text{future}_{(u_n, 0)}(1) = \{u_1\}$. Then, at time 2, $u_n$ meets $u_2$ and, by Lemma 1, $u_1$ meets $u_3$ via the edge than is perpendicular to $\{u_n, u_2\}$, thus $\text{future}_{(u_n, 0)}(2) = \{u_1, u_2, u_3\}$. We show that for all times $t$ it holds that $\text{future}_{(u_n, 0)}(t) = \{u_1, \ldots, u_{2t-1}\}$. The base case is true since $\text{future}_{(u_n, 0)}(1) = \{u_1\}$. It is not hard to see that, for $t \geq 2$, $N_{u_2}(t) = 2t - 2$, $N_{u_1}(t) = 2t - 1$, and for all $u_i \in \text{future}_{(u_n, 0)}(t) \setminus \{u_1, u_2\}$, $1 \leq N_{u_i}(t) \leq 2t - 2$. Now consider time $t + 1$. Lemma 1 gurantees now that for all $u_i \in \text{future}_{(u_n, 0)}(t)$ we have that $N_{u_i}(t + 1) = N_{u_i}(t) + 2$. Thus, the only new influences at step $t + 1$ are by $u_1$ and $u_2$ implying that $\text{future}_{(u_n, 0)}(t + 1) = \{u_1, \ldots, u_{2(t+1)-1}\}$. Consequently, the **oit** is 1.

The edge-period is $n - 1$ because the edges leaving the center appear one after the other in a clockwise fashion, thus taking $n - 1$ steps to any such edge to reappear, and, by construction, any other edge appears only when its unique perpendicular that is incident to the center appears (thus, again every $n - 1$ steps). □

It is interesting to note that in dynamic graphs with a static set of nodes, if at least one change happens each time, then every instance $G(t)$ will eventually reappear after at most $\sum_{k=0}^{\binom{n}{2}} \binom{\binom{n}{2}}{k}$ steps. This counts all possible different graphs of $n$ vertices with $k$ edges and sums for all $k \geq 0$. Thus the edge-period is bounded from above by a function of $n$. Note also that Theorem 1 is optimal w.r.t. edge-periodicity as *it is impossible to achieve at the same time unit* **oit** *and edge-period strictly greater than* $n - 1$. To see this, notice that if no edge is allowed to reappear in less than $n$ steps then any node must have no neighbors once every $n$ steps.

---

[3] A *factor* of a graph is a spanning subgraph: a subgraph whose vertex set is the whole graph. More on factors can be found in [MR85].

### 4.3 Some Further Interesting Properties

Here we show that one cannot guarantee at the same time unit oit and at most one outgoing influence per node per step. In fact, we conjecture that unit oit implies that some node disseminates in $\lfloor n/2 \rfloor$ steps.

Consider the following influence metric:

**Definition 4.** *Define the* maximum outgoing influence *(moi) of a dynamic graph $G = (V, E)$ as the maximum $k$ for which $\exists u \in V$ and $\exists t, t' \in \mathbb{N}$, $t' \geq t$, s.t. $|\text{future}_{(u,t)}(t'+1)| - |\text{future}_{(u,t)}(t')| = k$.*

In words, the moi of a dynamic graph is the maximum number of nodes that are ever concurrently influenced by the time-state of a node. We now prove an interesting theorem stating that if one tries to guarantee unit oit then she must necessarily accept that at some steps more than one outgoing influences of the same time-node will occur leading to faster dissemination than $n - 1$ for this particular node.

**Theorem 2.** *The moi of any dynamic graph with $n \geq 3$ and unit oit is at least 2.*

*Proof.* For $n = 3$, just notice that unit oit implies that, at any time $t$, some node has necessarily 2 neighbors. We therefore focus on $n \geq 4$. For the sake of contradiction, assume that the statement is not true. Then at any time $t$ any node $u$ is connected to exactly one other node $v$ (at least one neighbor is required for oit 1 - see Proposition 1 - and at most one is implied by our assumption). Unit oit implies that, at time $t + 1$, at least one of $u, v$ must be connected to some $w \in V \backslash \{u, v\}$, let it be $v$. Proposition 1 requires that also $u$ must have an edge labeled $t + 1$ incident to it. If that edge arrives at $v$, then $v$ has 2 edges labeled $t + 1$. If it arrives at $w$, then $w$ has 2 edges labeled $t + 1$. So it must arrive at some $z \in V \backslash \{u, v, w\}$. Note now that, in this case, the $(t-1)$-state of $u$ first influences both $w, z$ at time $t + 1$ which is contradictory, consequently the moi must be at least 2. □

In fact, notice that the above theorem proves something stronger: Every second step at least half of the nodes influence at least 2 new nodes each. This, together with the fact that it seems to hold for some basic cases, makes us suspect that the following conjecture might be true:

*Conjecture 1.* If the oit of a dynamic graph is 1 then $\forall t \in \mathbb{N}$, $\exists u \in V$ s.t. $|\text{future}_{(u,t)}(t + \lfloor n/2 \rfloor)| = n$.

That is, if the oit is 1 then, in every $\lfloor n/2 \rfloor$-window, some node influences all other nodes (e.g. influencing 2 new nodes per step).

### 4.4 Termination and Computation

We provide here a termination criterion by answering the following question: "Given that the oit of a dynamic graph is $k$ what is its iit?" If a node knows the iit then it can use it as an upper bound on the time needed to hear from a new node. If this time ellapses without a new influence then it knows that it has already been influenced by all nodes. We then establish a computational reduction to 1-interval connected graphs.

**Theorem 3.** *In a dynamic graph with oit $k$, take a node $u$ and a time $t$ and denote $|\text{past}_{(u,t)}(0)|$ by $l$. It holds that $|\text{past}_{(u,t+\lceil kl(l+1)/2\rceil-1)}(0)| \geq \min\{l + 1, n\}$.*

*Proof.* Consider a node $u$ and a time $t$ and define $A_u(t) := \text{past}_{(u,t)}(0)$ (we only prove it for the initial states of nodes but easily generalizes to any time), $I_u(t') := \{v \in A_u(t) : A_v(t') \backslash A_u(t) \neq \emptyset\}$, $t' \geq t$, that is $I_u(t')$ contains all nodes in $A_u(t)$ whose $t'$-states have been influence by nodes not in $A_u(t)$ (these nodes know new info for $u$), $D_u(t') := A_u(t) \backslash I_u(t')$, that is all nodes in $A_u(t)$ that do not know new info, and $l := |A_u(t)|$. The only interesting case is for $V \backslash A_u(t) \neq \emptyset$. Since the oit is k we have that at most by round $t + kl$, $(u,t)$ influences some node in $V \backslash D_u(t)$ say via some $u_2 \in D_u(t)$. By that time, $u_2$ leaves $D_u$. Next consider $(u, t+kl+1)$. In $k(l-1)$ steps it must influence some node in $V \backslash D_u$ since now $u_2$ is not in $D_u$. Thus, at most by round $t + kl + k(l-1)$ another node, say e.g. $u_3$, leaves $D_u$. In general, it holds that $D_u(t' + k|D_u(t')|) \leq \max\{|D_u(t')| - 1, 0\}$. It is not hard to see that at most by round $j = t + k(\sum_{1 \leq i \leq l} i) - 1$, $D_u$ becomes empty, which by definition implies that $u$ has been influenced by the initial state of a new node. In summary, $u$ is influenced by another initial state in at most $k(\sum_{1 \leq i \leq l} i) - 1 = [kl(l+1)/2] - 1$ steps. □

We next show that there is indeed a dynamic graph with iit much greater than oit.

**Theorem 4.** *There is a dynamic graph with oit $k$ but iit $k(n-3)$.*

*Proof.* Consider the dynamic graph $G = (V, E)$ s.t. $V = \{u_1, u_2, \ldots, u_n\}$ and $u_i$, for $i \in \{1, n-1\}$, is connected to $u_{i+1}$ via edges labeled $jk$ for $j \in \mathbb{N}_{\geq 1}$, $u_i$, for $i \in \{2, 3, \ldots, n-2\}$, is connected to $u_{i+1}$ via edges labeled $jk$ for $j \in \mathbb{N}_{\geq 2}$. and $u_2$ is connected to $u_i$, for $i \in \{3, \ldots, n-1\}$ via edges labeled $k$. In words, at step $k$, $u_1$ is only connected to $u_2$, $u_2$ is connected to all nodes except from $u_n$ and $u_n$ is connected to $u_{n-1}$. Then every multiple of $k$ there is a single linear graph starting from $u_1$ and ending at $u_n$. At step $k$, $u_2$ is influenced by the initial states of nodes $\{u_3, \ldots, u_{n-1}\}$. Then at step $2k$ it forwards these influences to $u_1$. Since there are no further shortcuts, $u_n$'s state needs $k(n-1)$ steps to arrive at $u_1$, thus there is an incoming-influence-gap of $k(n-2)$ steps at $u_1$. To see that oit is indeed $k$ we argue as follows. Node $u_1$ cannot use the shortcuts, thus by using just the linear graph it influences a new node every $k$ steps. $u_2$ influences all nodes apart from $u_n$ at time $k$ and then at time $2k$ it also influences $u_n$. All other nodes do a shortcut to $u_2$ at time $k$ and then for all multiples of $k$ their influences propagate to both directions from two sources: theirself and $u_2$ influencing 1 to 4 new nodes every $k$ steps. □

We now show that the $[kl(l+1)/2] - 1$ ($l := |\text{past}_{(u,t)}(0)|$) upper bound on the iit is optimal in the following sense: a node cannot obtain a better upper bound based solely on $l$.

**Theorem 5.** *Assume that a node $u$ computes an $f(l)$ based on $l := |\text{past}_{(u,t)}(0)|$ and if it does not get a new influence by time $f(l)$ outputs $l$ and halts. If $f(l) = o(kl^2)$ then the count may be incorrect.*

*Proof.* Consider the set $\text{past}_{(u,t)}(0)$. Take any dynamic subgraph on $\text{past}_{(u,t)}(0)$, disconnected from the rest of the nodes, that satisfies oit $= k$ and that all nodes in $\text{past}_{(u,t)}(0)$ need $\Theta(kl)$ rounds to causally influence all other nodes in $\text{past}_{(u,t)}(0)$; this could, for example, be the alternating matchings graph from Section 4.2 with one matching appearing in rounds that are odd multiples of $k$ and the other in even. In $\Theta(kl)$ rounds, say in round $j$, some *intermediary* node $v \in \text{past}_{(u,t)}(0)$ must get the outgoing influences of nodes in $\text{past}_{(u,t)}(0)$ outside $\text{past}_{(u,t)}(0)$ so that they continue to influence new nodes. Assume that in round $j-1$ the adversary directly connects all nodes in $\text{past}_{(u,t)}(0) \backslash \{v\}$ to $v$. In this way, at time $j$, $v$ forwards outside $\text{past}_{(u,t)}(0)$ the $(j-2)$-states (and all previous ones) of all nodes in $\text{past}_{(u,t)}(0)$. Provided that $V \backslash \text{past}_{(u,t)}(0)$ is sufficiently big (that is to say $\Omega(kl)$), the adversary can now keep $D = \text{past}_{(u,t)}(0) \backslash \{v\}$ disconnected from the rest of the nodes for another $\Omega(kl)$ rounds (in fact, one round less this time) without violating oit $= k$ as the new influences of the $(j-2)$-states of nodes in $D$ may keep occuring outside $D$. The same process repeats by a new *intermediary* $v_2 \in D$ playing the role of $v$ this time. Each time the process repeats, in $\Theta(|D|)$ rounds the intermediary gets all outgoing influences outside $D$ and is then removed from $D$. It is straightforward to observe that a new incoming influence needs $\Omega(kl^2)$ rounds to arrive at $u$ in such a dynamic network. □

However, note that the above negative result does not exclude the possibility that some protocol uses more involved info for a node to determine whether it has heard from all nodes in $V$.

Now we associate the oit and iit metrics of a dynamic graph with 1-interval connectivity. In particular, Corollary 1 establishes that protocols that are correct in 1-interval graphs carry over to graphs with oit or iit $k$ with only a $O(kn)$ delay being introduced in the process.

**Theorem 6.** *Assume that the oit or the iit of a dynamic graph, $G = (V, E)$, is $k$. Then for all times $t$ the graph $(V, \bigcup_{i=t+1}^{t+k\lfloor n/2 \rfloor} E(i))$ is connected.*

*Proof.* It suffices to show that for any partitioning $(V_1, V_2)$ of $V$ there is an edge in the cut labeled from $\{t+1, \ldots, t+k\lfloor n/2 \rfloor\}$. W.l.o.g. let $V_1$ be the smallest, thus $|V_1| \leq \lfloor n/2 \rfloor$. Take any $u \in V_1$. By definition of oit, $|\text{future}_{(u,t)}(t+k\lfloor n/2 \rfloor)| \geq |\text{future}_{(u,t)}(t+k|V_1|)| \geq |V_1| + 1$ implying that some edge in the cut has transferred $u$'s $t$-state out of $V_1$ at some time in $t+1, \ldots, t+k\lfloor n/2 \rfloor$. The proof for iit=$k$ is similar. □

**Corollary 1.** *Any $f(n)$-time protocol that is correct on 1-interval-connected graphs has an equivalent $k\lfloor n/2 \rfloor f(n)$-time protocol for graphs with either oit or iit equal to $k$.*

*Proof.* The dynamic graph $G' = (V, E')$, where $E'(t) = \bigcup_{i=(t-1)k\lfloor n/2 \rfloor+1}^{tk\lfloor n/2 \rfloor} E(i)$, $t \geq 1$, is 1-interval connected. □

Now we define a more natural and practical metric for capturing the temporal connectivity of a possibly disconnected dynamic network the we call the *connectivity time* (ct).

**Definition 5.** *We define the* connectivity time *(ct) of a dynamic network $G = (V, E)$ as the minimum $k \in \mathbb{N}$ s.t. for all times $t \in \mathbb{N}$ the static graph $(V, \bigcup_{i=t}^{t+k-1})$ is connected.*

In words, the ct of a dynamic network is the maximal time of keeping any two parts of the network disconnected. That is to say, in every ct-window of the network an edge appears in every $(V_1, V_2)$-cut.

**Proposition 2.** oit $\leq$ ct.

*Proof.* We show that for all $u \in V$ and all times $t, t' \in \mathbb{N}$ s.t. $t' \geq t$ it holds that $|\text{future}_{(u,t)}(t'+\text{ct})| \geq \min\{|\text{future}_{(u,t)}(t')| + 1, n\}$. Assume $V \setminus \text{future}_{(u,t)}(t') \neq \emptyset$ (as the other case is trivial). In at most ct rounds an edge joins $\text{future}_{(u,t)}(t')$ to $V \setminus \text{future}_{(u,t)}(t')$. Thus, in at most ct rounds $\text{future}_{(u,t)}(t')$ increases by at least one. □

We now show that the above does not hold with equality and in fact it may be violated in the worst possible way.

**Proposition 3.** *There is a dynamic graph with unit oit but ct $= \Omega(n)$.*

*Proof.* Recall the alternating matchings on a ring dynamic graph from Section 4.2. Now take any set $V$ of a number of nodes that is a multiple of 4 (this is just for simplicity and is not necessary) and partition it into two sets $V_1, V_2$ s.t. $|V_1| = |V_2| = n/2$. If each part is an alternating matchings graph for $|V_1|/2$ rounds then every $u$ say in $V_1$ influences 2 new nodes in each round and similarly for $V_2$. Clearly we can keep $V_1$ disconnected from $V_2$ for $n/4$ rounds without violating the fact that oit $= 1$. □

Now assume that all nodes know some upper bound $T$ on the ct.

**Theorem 7.** *Node $u$ knows at time $t$ that $\text{past}_{(u,t)}(0) = V$ iff $\text{past}_{(u,t)}(0) = \text{past}_{(u,t)}(T)$.*

*Proof.* If $\text{past}_{(u,t)}(0) = \text{past}_{(u,t)}(T)$ then we have that $\text{past}_{(u,t)}(T) = V$. The reason is that $|\text{past}_{(u,t)}(0)| \geq \min\{|\text{past}_{(u,t)}(T)| + 1, n\}$. To see this, assume that $V \backslash \text{past}_{(u,t)}(T) \neq \emptyset$. At most by round $T$ there is some edge joining some $w \in V \backslash \text{past}_{(u,t)}(T)$ to some $v \in \text{past}_{(u,t)}(T)$. Thus, $(w, 0) \leadsto (v, T) \leadsto (u, t) \Rightarrow w \in \text{past}_{(u,t)}(0)$. In words, all nodes in $\text{past}_{(u,t)}(T)$ belong to $\text{past}_{(u,t)}(0)$ and at least one node not in $\text{past}_{(u,t)}(T)$ (if one exists) must belong to $\text{past}_{(u,t)}(0)$.

For the other direction, assume that there exists $v \in \text{past}_{(u,t)}(0) \backslash \text{past}_{(u,t)}(T)$. This does not imply that $\text{past}_{(u,t)}(0) \neq V$ but it does imply that even if $\text{past}_{(u,t)}(0) = V$ node $u$ cannot know it has heard from everyone. Note that $u$ heard from $v$ at some time $T' < T$ but has not heard from $v$ since then. It can be the case that arbitrarily many nodes were connected to no node until time $T - 1$ and from time $T$ onwards were connected only to node $v$ ($v$ in some sense conceals these nodes from $u$). As $u$ has not heard from the $T$-state of $v$ it can be the case that it has not heard at all from arbitrarily many nodes, thus it cannot decide on the count. □

## 5 Local Communication Windows

We assume here an underlying *communication network*, which is modeled by an undirected simple connected graph $C = (V, A)$, where $V$ is a set of nodes and $A$ is a set of undirected edges. We associate each $u \in V$ with a finite integer $c_u \geq 1$, called $u$'s *neighborhood cover time* (or *cover time* for simplicity), and let $c \in \mathbb{N}^V$ be a vector of cover times indexed by nodes. We denote by $N(u)$ the *neighborhood* of node $u$ in $C$, that is $N(u) = \{v \in V : \{u, v\} \in A\}$.

A dynamic graph $G = (V, E)$ has now $E : \mathbb{N}_{\geq 1} \to \mathcal{P}(A)$. We say that $G$ *respects* a vector of cover times $c \in \mathbb{N}^V$ if for all $r \in \mathbb{N}$ and all $u \in V$ it holds that $\{v \in V : \{u, v\} \in \bigcup_{i=r}^{r+c_u-1} E(i)\} = N(u)$ (or $\geq$ in case we would allow a node to possibly communicate outside its underlying neighborhood); that is each node $u$ must cover all its possible neighbors in any $c_u$-window of the dynamic graph. Note that again we do not require the instantaneous graphs to be connected. Nonetheless, it is not hard to see that this definition guarantees that each node may eventually influence every other node in the network. We are interested in protocols that are correct for all possible pairs $(G, c)$, where $G$ is a dynamic graph that respects the vector of cover times $c$.

First note that if nodes do not know their cover times nor some upper bound on them, then non-trivial halting computations are impossible. To see this, consider any protocol that terminates in $k$ steps on some dynamic graph $G$. Now augment $G$ with some dynamic graph $D$ that has its first communication with $G$ at time $k + 1$ and notice that termination on $G$ occurs at time $k$ without any consideration of $D$.

We focus on the case in which each node $u$ knows its precise cover time $c_u$. First of all, notice that for all cover times $c$ there is a dynamic graph that respects $c$, namely the static graph in which $E(r) = A$ for all $r \in \mathbb{N}$. However, not all cover times $c$ *admit a worst-case dynamic graph*, that is one in which for all $u \in V$ there is an $r \in \mathbb{N}$ such that $|\{v \in V : \{u, v\} \in \bigcup_{i=r}^{r+c_u-2} E(i)\}| < |N(u)|$. It is not hard to see that a cover-time vector $c$ admits a worst-case dynamic graph $G$ iff $\forall u \in V, \exists v \in N(u)$ such that $c_v \geq c_u$.

An interesting question is whether nodes can verify if a given vector of cover times admits a worst-case dynamic graph. In fact, we want nodes to accept if all cover times are consistent and fix inconsistent cover times otherwise. Let $C_u$ be an upper bound on $c_u$. Each node $u$ must check whether there is some $v \in N(u)$ such that $C_v \geq C_u$. $u$ broadcasts $C_u$ for $C_u$ rounds. If $C_v < C_u$ for all $v \in N(u)$, then $u$ sets $C_u$ to $\max_{v \in N(u)}\{C_v\}$, otherwise it accepts.

We now deal with the problem of information dissemination and counting and present a protocol for the latter problem.

Let $u = u_0, u_1, \ldots, u_k = v$ be a simple path $p$ joining nodes $u, v$. The worst case time for $u$ to influence $v$ by messages traversing $p$ is $l(p) = \sum_{i=1}^{k-1} \min\{c_{u_i}, c_{u_{i+1}}\}$ (called *length* or *maximum delay*). Extend $l$ as $l(u,v) = \min_{p \in P(u,v)} l(p)$, where $P(u,v)$ is the set of all simple paths joining $u, v$. In the dynamic networks under consideration we have that the dynamic diameter is $D = \max_{u,v \in V} l(u,v)$. It is obvious that if all nodes knew some upper bound $H \geq D$ then each node could halt after $H$ rounds knowing that it has influenced and been influenced by all other nodes. A natural question is whether nodes can achieve this without knowing $D$ in advance. For example, is there a terminating algorithm for *counting* (i.e. for computing $n$) if nodes only know their exact cover times? In the sequel, we answer this question in the affirmative.

Let

- $\mathrm{psum}_{(u,t')}(t) := \sum_{v \in \mathrm{past}_{(u,t')}(t)} c_v$, and
- $\mathrm{fsum}_{(u,t)}(t') := \sum_{v \in \mathrm{future}_{(u,t)}(t')} c_v$.

**Lemma 2.** *For all times $t, t'$ such that $t \leq t'$, all nodes $u, v$, and all $k \geq 1$, if $v \in \mathrm{past}_{(u,t')}(t)$ for all $E$ then $v \in \mathrm{past}_{(u,t'+k)}(t+k)$.*

*Proof.* Take any $v \in \mathrm{past}_{(u,t')}(t)$. To show that $v \in \mathrm{past}_{(u,t'+k)}(t+k)$ we notice that for any dynamic edge function $E'$ there exists $E$ such that $E'(r+k) = E(r)$ for all $t \leq r \leq t'$. □

**Lemma 3.** $\mathrm{past}_{(u,t)}(0) \subseteq \mathrm{past}_{(u,\mathrm{psum}_{(u,t)}(0)-c_v)}(0)$.

*Proof.* We show that $v \in \mathrm{past}_{(u,t)}(0)$ implies $v \in \mathrm{past}_{(u,\mathrm{psum}_{(u,t)}(0)-c_v)}(0)$. The time-node $(v,0)$ has influenced $(u,t)$ via a simple path $p$ that only visits nodes from $\mathrm{past}_{(u,t)}(0)$ since $(v,0) \rightsquigarrow (w,j) \rightsquigarrow (u,t)$ for any intermediate node $w$ implies $w \in \mathrm{past}_{(u,t)}(0)$; to see this note that $(w,0) \rightsquigarrow (w,j)$ for all $j \geq 0$. Clearly, the longest such path $p'$ is a path that is hamiltonian in $C[\mathrm{past}_{(u,t)}(0)]$ [4] beginning from $u$ and ending at $v$. Since $l(p) \leq l(p') \leq \mathrm{psum}_{(u,t)}(0) - c_v$ and $(v,0) \rightsquigarrow (u, l(p))$ it also holds that $(v,0) \rightsquigarrow (u, \mathrm{psum}_{(u,t)}(0) - c_v)$ or equivalently $v \in \mathrm{past}_{(u,\mathrm{psum}_{(u,t)}(0)-c_v)}(0)$. □

**Lemma 4.** *For all nodes $u \in V$ and times $t \geq 0$ we have:*

1. $|\mathrm{past}_{(u,\mathrm{psum}_{(u,t)}(0))}(0)| \geq \min\{|\mathrm{past}_{(u,t)}(0)| + 1, n\}$ *and*
2. $|\mathrm{future}_{(u,0)}(\mathrm{fsum}_{(u,0)}(t))| \geq \min\{|\mathrm{future}_{(u,0)}(t)| + 1, n\}$.

*Proof.* We only prove the first statement since the second is symmetric. The only interesting case is when $|\mathrm{past}_{(u,t)}(0)| < n$ in which case there exists $w \in V \setminus \mathrm{past}_{(u,t)}(0)$. By Lemma 3, $\mathrm{past}_{(u,t)}(0) \subseteq \mathrm{past}_{(u,\mathrm{psum}_{(u,t)}(0)-c_v)}(0) \subseteq \mathrm{past}_{(u,\mathrm{psum}_{(u,t)}(0))}(0)$. So we just need to show that there is a $w \in \mathrm{past}_{(u,\mathrm{psum}_{(u,t)}(0))}(0) \setminus \mathrm{past}_{(u,t)}(0)$. Connectivity ensures that there is some $\{w,v\} \in A$, for $w \in V \setminus \mathrm{past}_{(u,t)}(0)$ and $v \in \mathrm{past}_{(u,t)}(0)$. Clearly $(w,0) \rightsquigarrow (v, c_v)$. Since $(v,0) \rightsquigarrow (u, \mathrm{psum}_{(u,t)}(0) - c_v)$, by Lemma 2 $(v, c_v) \rightsquigarrow (u, \mathrm{psum}_{(u,t)}(0))$. Transitivity ensures that $(w,0) \rightsquigarrow (u, \mathrm{psum}_{(u,t)}(0))$ and $w \in \mathrm{past}_{(u,\mathrm{psum}_{(u,t)}(0))}(0)$. □

Lemma 4 provides us with the following criterion for a node to detect when it has been causally influenced by all other nodes: $|\mathrm{past}_{(u,\mathrm{psum}_{(u,t)}(0))}(0)| = |\mathrm{past}_{(u,t)}(0)| \Rightarrow |\mathrm{past}_{(u,t)}(0)| = V$. That is, at any time $t$, any new influence of the state of $u$ by some initial state must occur at most by time $\mathrm{psum}_{(u,t)}(0)$. If this time elapses without any new influence then $u$ knows that it has been causally influenced by all other nodes. An easier to perform but equivalent test is the following: $t = \text{psum}_{(u,t)}(0) \Rightarrow |\text{past}_{(u,\text{psum}_{(u,t)}(0))}(0)| = |\text{past}_{(u,t)}(0)| \Rightarrow |\text{past}_{(u,t)}(0)| = V$. In the following proposition, we use the latter criterion to solve counting.

---
[4] We denote by $G[V']$ the subgraph of a graph $G$ induced by nodes in $V'$.

But first define an edge weight $w(e)$ for each edge $e \in A$ as $w(e) := min c_u, c_v$. We are then guaranteed that an edge $e$ appears at least once in every time interval of length $w(e)$. This implies that within time $W := \sum_{e \in D(C)} w(e)$, where $D(C)$ is a diameter of $C$ (that is within time equal to the weighted diameter of $C$), everyone hears from everyone else and then another $\sum_{u \in V} c_u - W$ rounds are needed for the nodes to know that they are done.

**Proposition 4.** *Counting can be solved in $O(\sum_{u \in V} c_u)$ rounds using messages of size $O(n \log n + \sum_{u \in V} c_u)$.*

*Proof.* Each node $u$ maintains a set of UIDs $A_u$, where initially $A_u(0) = \{u\}$, and a vector $c_u[]$ of cover times indexed by UIDS in $A_u$, where initially $c_u = (c_u)$. In each round $r$, $u$ sends $(A_u, c_u)$ to all its current neighbors, stores in $A_u$ all received UIDs and, for each new UID $v \notin A_u(r-1)$, $u$ stores $t_v$ in $c_u$. Moreover, nodes keep track of the round number. At the end of each round $r$, if $r = \sum_{v \in A_u(r)} c_u[v]$ node $u$ halts and outputs $|A_u|$; otherwise, $u$ continues on to the next round. □

## 6 Conclusions

We studied for the first time worst-case dynamic distributed systems that are only restricted by a minimal temporal connectivity condition. We proved that fast dissemination and computation are possible even under continuous disconnectivity. We also proposed a novel model of bounded local coverage for which we presented a protocol for counting.

There are many open problems and promising research directions related to this work. Is Conjecture 1 true? Namely, is it true that if the oit of a dynamic graph is 1 then there is always some node that only needs $\lfloor n/2 \rfloor$ steps to influence the whole network? Information dissemination is still only guaranteed under continuous broadcasting. How can the number of redundant transmissions be reduced? Is there a way to exploit visibility to this end? Does predictability help (i.e. some knowledge of the future)? Moreover, dynamic networks in which nodes have some partial control over their mobility have not been studied yet. All works so far in dynamic networks assume a unique sampling of the input and then a single execution on this input. However, there are natural settings in which such a unique sampling is impossible and the nodes have to consider the whole state of the environment. How can this be performed in a highly-dynamic setting in which also the environment may be dynamic? Finally, randomization will be certainly valuable in constructing fast and symmetry-free protocols. We strongly believe that these and other known open questions and research directions will motivate the further growth of this emerging field.

**Acknowledgements.** We would like to thank some anonymous reviewers for their very useful comments on a previous version of this work.